\documentclass[11pt, notitlepage]{article}
\usepackage{setspace}
\onehalfspace

\usepackage[utf8]{inputenc}
\usepackage{titling}
\usepackage[a4paper, total={6.5in, 8.5in}]{geometry}
\usepackage{algpseudocode}
\usepackage{setspace}
\usepackage{etoc}
\usepackage{etoolbox}
\AtBeginEnvironment{algorithmic}{\setstretch{1.8}}

\usepackage{amsmath}

\usepackage{amsthm}

\usepackage{xcolor}
\usepackage[allbordercolors=white]{hyperref}

\usepackage{graphicx}
\usepackage{authblk}
\title{Fundamental Challenges for On-Chip Diffractive Processing at Gigahertz Speeds}
\author[1,*]{Benjamin Wetherfield}
\author[1]{Timothy D. Wilkinson}
\affil[1]{Electrical Engineering Division, Department of Engineering, University of Cambridge}
\affil[*]{bsw28@cam.ac.uk}
\date{}

\begin{document}
\begin{titlingpage}
\maketitle
\begin{onehalfspace}
\begin{abstract}
The demands of proliferating big data and massive deep learning models, against a backdrop of a mounting climate emergency and the abating of Moore’s law, push technologists to develop high-speed, high-throughput, low energy and miniaturisable computer hardware. Using light as a fundamental resource, free-space optical computing and on-chip photonic computing devices provide, respectively, powers of natural parallelism and miniaturisability. Recent work harnessing diffractive effects inside planar (or slab) waveguides has seemingly combined the best elements of each competing technology. Yet, as this paper argues, certain challenges will emerge as the clock-speeds of on-chip diffractive systems are pushed to compete with legacy technologies. Using a ``time-aware'' analytical approach to wave propagation, a prediction is made of the presence of a time-based error term that has not yet been accounted for in on-chip diffractive architectures. System operating frequency bounds in the gigahertz range are quantified, above which time-based errors discernibly affect the accuracy of system performance, and below which the errors can safely be ignored, using design parameters from previously published  work. The analysis and related bounds presented hold value in any context where high throughput on-chip diffractive operations may be exploited, including beam-shaping, spectroscopy, sensing and communications.
\end{abstract}
\end{onehalfspace}
\end{titlingpage}

  As massive deep learning software stacks continue to grow and Moore's law, governing computer hardware capacities, abates, interest in optical and photonic computing platforms is growing to meet the demands of modern challenges \cite{liChallengesModernComputing2021}.
  Methods encompassing continuous-wave on-chip photonic techniques (as well as synapse-inspired neuromorphic photonic designs) and free-space optical methods have been shown capable of implementing a wide range of artificial intelligence and neural network infrastructures with light-speed processing \cite{shastriPhotonicsArtificialIntelligence2021}. Until recently, technical approaches involving continuous-wave light sources on-chip and in free-space have seen little overlap, with propagation of light being harnessed in one and three free dimensions respectively. Here, diffractive on-chip photonics \cite{zareiIntegratedPhotonicNeural2020,ongPhotonicConvolutionalNeural2020,fuOnchipPhotonicDiffractive2021,yanAllopticalGraphRepresentation2022a,zhuSpaceefficientOpticalComputing2022,fuPhotonicMachineLearning2023,levyInhomogenousDielectricMetamaterials2007,liaoAIassistedOnchipNanophotonic2020} presents an exciting new paradigm that bridges the gap, harnessing \emph{two} dimensional (2D) wave propagation and taking the best elements from each parent field, with the natural parallelism of free-space optics meeting the miniaturisability and fine, per-channel control of photonic methods. As we show in this paper, however, modelling techniques used until now have not done sufficient justice to the true nature of 2D wave propagation. Wavefronts ``spread out'' over time in 2D, which qualitatively differs from the behaviour of wave propagation in 1D or 3D, where, by contrast, ``Huygens' principle'' holds \cite{gburMathematicalMethodsOptical2011}. Through ``time-aware'' mathematical modelling and analysis, we show that there are time-based error terms that have yet to be recognised in analysing and optimising diffractive on-chip systems. We demonstrate that these errors gain in significance as the operating frequencies of diffractive processing systems increase. As such, we argue that if diffraction on-chip is to process data at the rates required of modern applications -- not only in computing, but also in communications, sensing, beam-shaping and beyond -- great care will be needed to compensate for temporal errors, and to extract the most value possible from this promising technology.

  Before being ``flattened'' onto photonic chips, diffraction in free-space has already been shown to give rise to diverse parallelised computational resources, passively and at the speed of light. The parallelism inherent in optical diffraction emerges in several ways. First, given a wavefront consisting of a large plane of inputs, a Fourier transform can be approximated without conventional computation, using optics as simple as a single lens \cite{goodmanIntroductionFourierOptics2005}. This facilitates the performance of large convolutions at scale, through the insertion of metasurfaces \cite{silvaPerformingMathematicalOperations2014} or dynamic spatial light modulators \cite{changHybridOpticalelectronicConvolutional2018,yanFourierspaceDiffractiveDeep2019} in the Fourier domain, enabling efficient implementations of convolutional neural networks \cite{changHybridOpticalelectronicConvolutional2018,yanFourierspaceDiffractiveDeep2019}. Wave propagation itself, meanwhile, provides a many-to-many mapping from input pixels to output pixels, which has been harnessed to great effect in implementations of the inter-layer weight functions of convolutional neural networks \cite{linAllopticalMachineLearning2018,yanFourierspaceDiffractiveDeep2019,zhouLargescaleNeuromorphicOptoelectronic2021}. Beyond diffractive deep neural networks (D2NN's), the scalable parallelism of free-space optics shows immense versatility, with applications encompassing mode conversion systems \cite{fontaineLaguerreGaussianModeSorter2019}, arbitrary linear transforming devices \cite{spallFullyReconfigurableCoherent2020a,kulceAllopticalSynthesisArbitrary2021}, optical Fourier transform devices \cite{macfadenOpticalFourierTransform2017}, reservoir computing devices \cite{buenoReinforcementLearningLargescale2018}, extreme learning machines \cite{pierangeliPhotonicExtremeLearning2021} and optical Ising machines \cite{pierangeliLargescalePhotonicIsing2019}.

  In contrast to free-space methods, on-chip photonic techniques have built on the availability of mature silicon chip foundry methods and are readily miniaturisable to the scales expected of modern computer chips. Moreover, work by Miller has improved the practical up-take of such devices, through easily configurable designs with imperfect optical components that can be progressively aligned \cite{millerSelfaligningUniversalBeam2013,millerPerfectOpticsImperfect2015,millerSettingMeshesInterferometers2017}. On the other hand, parallelism does not come for free in these devices, and the number of components needed scale at least as $N^{2}$ in the number of inputs, for standard arrangements using rectangular waveguides, (thermo-optic) phase modulators (TOPMs), Mach-Zehnder interferometers (MZIs) and multimode interference couplers. That said, using these building blocks, compact implementations of unitary transforms \cite{reckExperimentalRealizationAny1994,clementsOptimalDesignUniversal2016} and, thereafter, arbitrary linear transforming devices, have been demonstrated \cite{millerSelfconfiguringUniversalLinear2013, ribeiroDemonstration4portUniversal2016,annoniUnscramblingLightAutomatically2017}, supporting applications to free-space modal beam separators \cite{milanizadehSeparatingArbitraryFreespace2022}, quantum information processing systems \cite{carolanUniversalLinearOptics2015,harrisQuantumTransportSimulations2017}, deep learning architectures \cite{shenDeepLearningCoherent2017,chengSiliconPhotonicsCodesign2020,zhangOpticalNeuralChip2021}, and wide-ranging re-programmable platforms \cite{bogaertsProgrammablePhotonicCircuits2020a}.

\begin{figure*}[t]
  \centering
\includegraphics[scale=0.8]{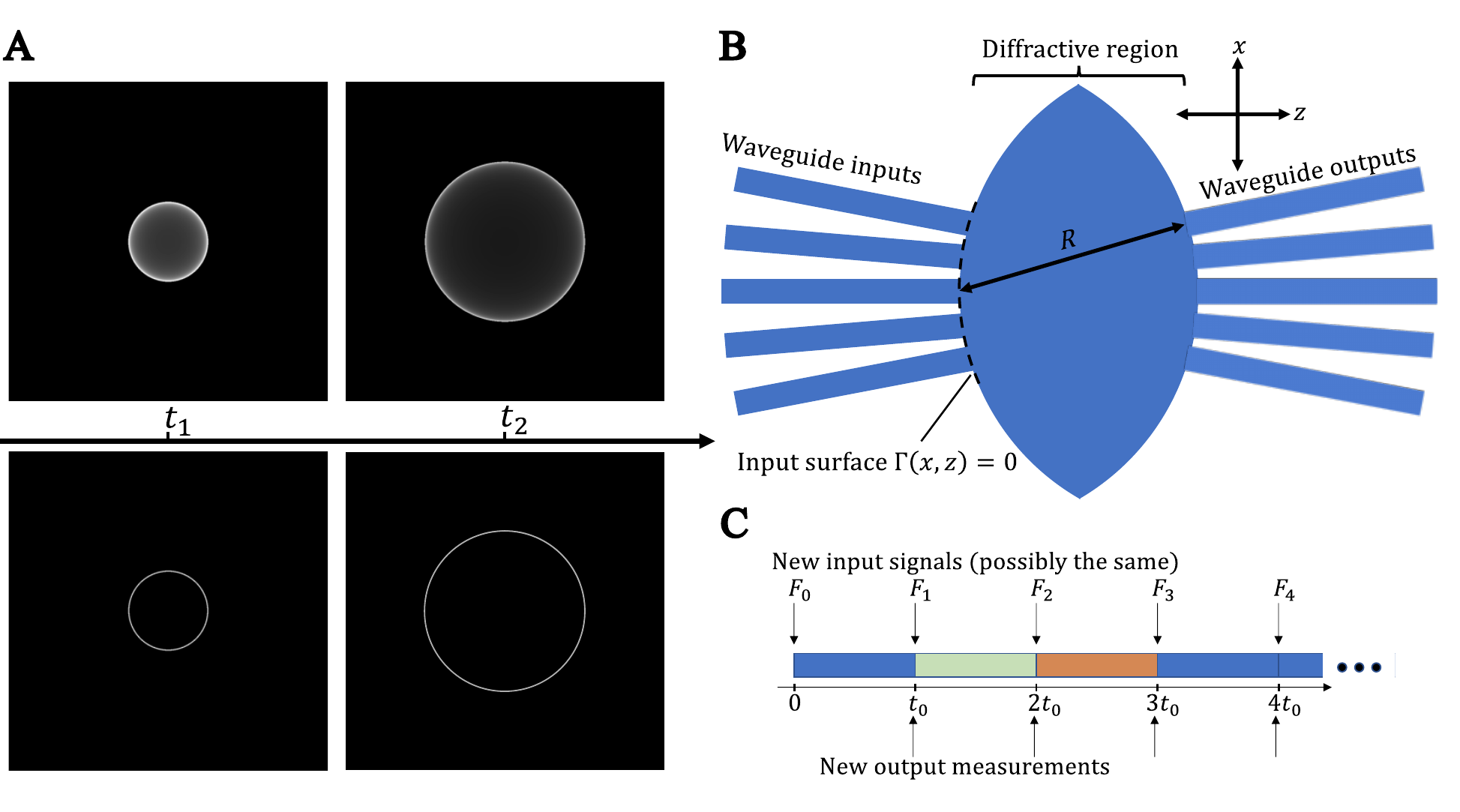}
  \caption{(A) Two snapshots of the propagation of an instantaneous point source in 2D (top) and 3D (bottom, planar cross-section) -- adapted from Gbur\protect \cite{gburMathematicalMethodsOptical2011}. (B) A star coupler architecture. The diffractive region (2D) features two circular arcs of radius $R$, separated by a distance $R$. The relationship between input and output signals can be approximated by a Fourier transform. (C) Diagram depicting the timeline of new inputs and outputs in the extended thought experiment.}
  \label{fig:compound}
\end{figure*}

  Recent work has proposed \cite{ongPhotonicConvolutionalNeural2020,zareiIntegratedPhotonicNeural2020,fuOnchipPhotonicDiffractive2021,yanAllopticalGraphRepresentation2022a} and experimentally demonstrated \cite{zhuSpaceefficientOpticalComputing2022,fuPhotonicMachineLearning2023} the possibility of combining the benefits of  diffractive and on-chip paradigms by integrating rectangular waveguide stages with planar (or slab) waveguide regions. The slab waveguide has one dimension constrained where a rectangular waveguide has two, leaving the remaining dimensions available for diffraction. Meanwhile, the architecture's waveguide format permits high efficiency coupling from rectangular waveguides.
  Star couplers, familiar from communications applications \cite{takiguchiOpticalOrthogonalFrequency2011,cincottiWhatElseCan2012} and depicted in Figure \ref{fig:compound}B, can be used to produce a discrete Fourier transform relationship between inputs and outputs, without the need for lens optics \cite{dragoneEfficientStarCouplers1989}. This enables the direct translation of many existing Fourier transform architectures from free-space optics \cite{changHybridOpticalelectronicConvolutional2018,yanFourierspaceDiffractiveDeep2019} directly into on-chip systems \cite{ongPhotonicConvolutionalNeural2020,zhuSpaceefficientOpticalComputing2022}.
Impressively, Zhu \emph{et al.} have demonstrated that star coupler-based computations involving around 6 times as much data can be performed with a nearly 2000-fold reduction in power consumption, compared to existing MZI-based photonic implementations, in a form factor one tenth the size \cite{zhuSpaceefficientOpticalComputing2022}. Their work concretely combines the benefits of free-space and conventional photonic paradigms by using TOPMs to modulate channels in the Fourier domain.
Further still, D2NN designs in free-space \cite{linAllopticalMachineLearning2018} have been adapted to on-chip settings, implementing neural network weights using 1D metasurfaces that are designed based on offline gradient descent optimisations, in dramatically reduced form factors \cite{zareiIntegratedPhotonicNeural2020,fuOnchipPhotonicDiffractive2021,fuPhotonicMachineLearning2023,yanAllopticalGraphRepresentation2022a};
in this vein, Fu \emph{et al.} fabricated a D2NN-type classification device that supports nearly 500 times the throughput of existing MZI-based photonic implementations \cite{fuPhotonicMachineLearning2023}.

In addition to miniaturising important architectures into chip-sized form factors, recent work has sought to borrow time-independent modelling tools from free-space in the analysis and optimisation of on-chip diffractive systems.
In layered diffractive architectures that adapt D2NNs, an angular spectrum \cite{yanAllopticalGraphRepresentation2022a} or Fresnel \cite{fuOnchipPhotonicDiffractive2021,fuPhotonicMachineLearning2023} model of diffraction has been used for modelling and for optimising the weights of 1D metasurfaces, in a manner analogous with free-space methods \cite{linAllopticalMachineLearning2018}.
Meanwhile, the validity of the star coupler as a Fourier transforming device, as well as a similar architecture supporting surface plasmonic polariton beam-shaping applications \cite{kouOnchipPhotonicFourier2016}, rests on the asymptotic approximation of a Hankel function by a complex exponential \cite{zhuSpaceefficientOpticalComputing2022}, using a small wavelength assumption that is standard in analogous approximations valid in free-space Fourier optics, such as the Fresnel and Fraunhofer formulations \cite{goodmanIntroductionFourierOptics2005}.

In this article, however, we argue that, for applications at high operating frequencies, the modelling assumptions from free-space need adjustment and compensation when descending into planar contexts, due to time-based effects associated with the diffusive nature of 2D wave propagation (illustrated in Figure \ref{fig:compound}A). Clues that new models are needed for high-speed systems may even be present in the experimental literature already. While important factors are accounted for, including fabrication imperfections, signal loading and signal detection errors, Fu \emph{et al.} report a 30\% difference in classification success between simulation and experiment that we hypothesise could, in part, result from temporal error, especially as their reported rapid data-rates rely on high-speed photodetection in the gigahertz range \cite{fuPhotonicMachineLearning2023}.

Hence, we adopt a ``time-aware'' modelling approach. Rather than using the time-independent models of free-space Fourier optics, we proceed from a time-\emph{dependent} driven wave equation. Constructing a suitable thought experiment, we are able to show that the outputs predicted by our model correspond to the predictions of time-independent methods with the addition of a time-based error. The error increases (by a factor of two) if the worst-case influence of previous inputs, which leave behind a diffusive residue, is taken into account. Each of the errors involves a special function, which we define as a form of ``incomplete Hankel function''. We derive bounded asymptotic approximations for this function, which allow us to bound the influence of time-based error on the accuracy of a system that is designed under the assumption of time-independence. Our final plot shows the derived frequency bounds visualised against previously published simulated and experimental work, allowing readers and researchers to weigh up the implications of our analysis for state-of-the-art photonic configurations.

\section*{Results}
\label{sec:analysis}

\subsection*{Time-Aware Modelling}
\label{sec:time-aware-modeling}

To distinguish the approach taken in this paper from previous work, we introduce some modelling distinctions. For a driven TE mode in a slab waveguide $\tilde{\boldsymbol{E}} = (E_{x}, E_{y}, E_{z})^{T}$, it can be shown that $E_{y}$ satisfies the following wave equation in the $x$ and $z$ (free) dimensions
\begin{equation}
  \label{eq:wave_eq_resscaled}
  \left (\frac{\partial^{2}}{\partial x^{2}} + \frac{\partial^{2} }{\partial z^{2}} - \frac{n_{e}^{2}}{c^{2}}\frac{\partial^{2} }{\partial t^{2}}\right)E_{y} = f(x,z,t)
\end{equation}
where $n_{e}$ is the effective refractive index of the mode in the medium, defined as $n_{e} = \beta/k$, where $\beta$ is the mode's propagation constant, $k$ is the vacuum wavenumber of the light and $f$ is the driving input field \cite{okamotoFundamentalsOpticalWaveguides2006}. $E_{x}$ and $E_{z}$ are then $0$ and the magnetic polarisation can be uniquely determined according to Maxwell's equations. The same holds with the roles of $\tilde{\boldsymbol{H}}$ and $\tilde{\boldsymbol{E}}$ reversed for TM modes. For notational consistency with other sources on Fourier optics \cite{goodmanIntroductionFourierOptics2005}, we will write $u$ in place of $E_{y}$ (or $H_{y}$, in the case of a TM mode).

In prior work, it has been tacitly or explicitly assumed that $u$ is separable in space and time, such that $u(x,z,t)~=~U(x,z)\exp(-i\omega t)$, with $\omega=kc$  and $U$ satisfies a homogeneous time-independent wave (or Helmoltz) equation with appropriate boundary conditions:
\begin{equation}
  \label{eq:driven_helmholtz}
  \left (\frac{\partial^{2}}{\partial x^{2}} + \frac{\partial^{2} }{\partial z^{2}} + \beta^{2}\right)U(x, z ) = 0
\end{equation}
which can be solved in terms of the following Green's function:
\begin{equation}
  \label{eq:3}
  G_{2D}(x,z) = -\frac{i}{4}H_{0}^{(1)}(\beta \rho)
\end{equation}
where $\rho = \sqrt{x^{2} + z^{2}}$, $H_{0}^{(1)}$ is a Hankel function of the first kind of order zero, and $G_{2D}$ solves the following equation:\cite{bakerMathematicalTheoryHuygens1950}
\begin{equation}
  \label{eq:G2D_solves}
  \left (\frac{\partial^{2}}{\partial x^{2}} + \frac{\partial^{2} }{\partial z^{2}} + \beta^{2}\right)G_{2D}(x-x', z-z') = \delta(x-x',z-z')
\end{equation}

But the assumption of the separability of $u$, while sufficient for certain low-operating-speed applications, is not true to the physical reality of 2D wave propagation. Diffraction in the plane, whether in the guise of slab waveguides, surface phenomena or even line sources in free-space, evinces diffusive behaviour over time, as illustrated in Figure \ref{fig:compound}A. Solving for $u$ at position $(x_{0}, z_{0})$ and time $t_{0}$ gives
\begin{equation}
  \label{eq:2d_driven}
  u(x_{0},z_{0},t_{0})=\frac{c_n}{2\pi} \iiint_{D} \frac{f(x,z, t)}{\sqrt{c_n^{2}(t_{0} - t)^{2} - \rho_{0}^{2}}} \; dx dz dt
\end{equation}
where $c_{n} = c/n_{e}$, $\rho_{0} = \sqrt{(x_{0}-x)^{2}+ (z_{0}-z)^{2}}$, and $D$ is defined by $0 \le t \le t_{0}$ and $\rho_{0}~\le~c_{n}(t_{0}-t)$ \cite{courantMethodsMathematicalPhysics1962}.

\subsection*{Setup: A Thought Experiment}
\label{sec:setup:-thought-exper}

In order to make use of our time-aware framing, we propose the following thought experiment. Suppose we have a system with inputs coupled from 1D waveguides into a 2D diffractive region. An example of one such setup, a star coupler, is shown in Figure \ref{fig:compound}B. The value of the input waveguides at the input surface $\Gamma(x,z)=0$ act as a source.

For our purposes, the driving function $f(x,z,t)$ encompasses the values of the inputs at the surface $\Gamma(x,z)=0$. Hence, assuming the system is started at time $0$, we may write:
\begin{equation}
  \label{eq:input}
  f(x,z,t) = \begin{cases} F(x,z) \exp(-i\omega t) \; &t \ge 0\\ 0 \; &t < 0
  \end{cases}
\end{equation}
where $F(x,z)$ encodes the relative complex amplitude as controlled inside each of the input waveguides and we set $F(x,z)=0$ formally for $\Gamma(x,z) \ne 0$.

\subsection*{Expression for Time-Based Error}
\label{sec:expr-time-based}

 Suppose that we take a measurement of the output of the system at a time $t_{0}$, where $t_{0}> \rho_{0}/c_{n}$ for all points of interest. Then if $(x_{0}, z_{0})$ denotes the position where the measurement is taken, we can plug (\ref{eq:input}) into (\ref{eq:2d_driven}) to give an expression for its value. Changing variables with $\tau=c_{n}(t_{0 } - t)/\rho_{0}$, we obtain:
\begin{align}
  u(x_{0}, z_{0}, t_{0}) = &\exp(-i\omega t_{0})\iint F(x,z) \nonumber \\
  &\times \int_{1}^{c_{n}t_{0}/\rho_{0}}-\frac{1}{2\pi}\frac{\exp(i \beta \rho_{0} \tau)}{\sqrt{\tau^{2} - 1}} \; d\tau \; dx dz
  \label{eq:change_vars_again}
\end{align}
where we can remove the leading phase term by absorbing it into the definition of $F(x,z)$. In the limit, $t_{0} \to \infty$ the integral in $\tau$ tends to $G_{2D}(x-x_{0},z-z_{0})$ (see page 170 of Watson's classic treatise \cite{watsonTreatiseTheoryBessel1966}). To examine the case with finite $t_{0}$, we define an incomplete Hankel function as follows:\footnote{Several definitions of incomplete Hankel functions have been proposed \protect \cite{agrestTheoryIncompleteCylindrical1971,cicchettiIncompleteHankelModified2004,jonesIncompleteBesselFunctions2007}. We follow closest the definition of ``Incomplete Cylindrical Functions of Poisson Form'' given in the monograph by Agrest and Maksimov \protect \cite{agrestTheoryIncompleteCylindrical1971}.}
\begin{equation}
  \label{eq:incomplete_hankel}
  H_{0}^{(1)}(\vert x \vert, T)=\frac{2}{i\pi}\int_{T}^{\infty}\frac{\exp(i \vert x \vert \tau)}{\sqrt{\tau^{2}-1}}\; d\tau
\end{equation}
setting $T\ge 1$, and which coincides with $H_{0}^{(1)}(\vert x \vert)$ when $T=1$. With this definition, we can write:
\begin{equation}
u(x_{0}, z_{0}, t_{0})= \iint F(x,z) G^{t_{0}}_{2D}(x_{0}-x,z_{0}-z) \; dx dz
\end{equation}
where $G^{t_{0}}_{2D}$ is a Green's function defined by
\begin{equation}
  \label{eq:Arrival_greens_2D}
  G^{t_{0}}_{2D}(x,z) = G_{2D}(x,z) + \frac{i}{4}H_{0}^{(1)}(\beta \rho, c_{n}t_{0}/\rho)
\end{equation}

The incomplete Hankel term can be thought of as an error from the ``ideal case'' suggested by time-independent methods of $G^{t_{0}}_{2D}\approx G_{2D}$.

\subsection*{Accounting for Previous Inputs}
\label{sec:acco-prev-inputs}

As well as the diffusive effects due to the current input, \emph{previous} inputs will leave a diffusive residue in the system, which will be picked up in the measurement of $u(x_{0}, z_{0}, t_{0})$. Suppose that the output is measured and the input is changed on a clock-tick of duration $t_{0}$ as shown in Figure \ref{fig:compound}C. At each tick, a new input is introduced to the system (with possible repeats). We wish to solve for the diffusive residue from previous inputs in the currently measured output.

We can sum the residue $R_{0}^{N-1}$ of the first $N$ inputs $\{F_{j}\}_{0}^{N-1}$ on the measurement of the output at time $(N+1)t_{0}$ as
\begin{align}
  &R_{0}^{N-1} (x_{0}, z_{0},(N+1)t_{0})
  =  \frac{i}{4}\sum_{j=1}^{N}\iint F_{N-j}(x,z) \nonumber \\
    &\times \left [H_{0}^{(1)}\left (\beta \rho_{0}, \frac{c_{n}}{\rho_{0}}(j+1)t_{0} \right) -H_{0}^{(1)}\left (\beta \rho_{0}, \frac{c_{n}}{\rho_{0}}jt_{0} \right)\right ] \; dx dz
    \label{eq:cumulative_effect_2}
\end{align}
The total residue is the result of interference from the error terms corresponding to each input $F_{j}$. In the worst case, interference between contributing terms of Equation (\ref{eq:cumulative_effect_2}) is totally constructive, which can occur when all of the previous inputs have had the same phase distribution. Adding the constraint that the total energy of each $F_{j}$ is fixed, we can assume, in the worst case,
 $F_{0} = F_{1} = \cdots$, yielding,
\begin{align}
  &R_{0}^{N-1}(x_{0}, z_{0},(N+1)t_{0})\bigg \rvert_{F_{0} = F_{1} = \cdots}   = \frac{i}{4}\iint F_{0}(x,z) \nonumber  \\
  &\times \left [H_{0}^{(1)}\left (\beta \rho_{0}, \frac{c_{n}}{\rho_{0}}(N+1)t_{0}\right ) -  H_{0}^{(1)}\left (\beta \rho_{0}, \frac{c_{n}}{\rho_{0}}t_{0}\right )\right] \; dx dz
\end{align}
which tends to the following bound for large $N$:
\begin{equation}
  \label{eq:large_N}
  -\frac{i}{4}\iint F_{0}(x,z)H_{0}^{(1)}\left (\beta \rho_{0}, \frac{c_{n}}{\rho_{0}}t_{0}\right ) \; dx dz
\end{equation}

Putting this result together with equation (\ref{eq:Arrival_greens_2D}), let us suppose that for arbitrarily large $N$, we have had $F_{j}(x,z) = -F(x,z)$ for some input function $F(x,z)$ and all $j < N$. Now, the $N$th input is flipped to $F_{N}(x,z)=F(x,z)$. The measured output on the next clock tick is given by
\begin{align}
  u(x_{0}, z_{0}, (N+1)t_{0}) &= \iint F(x,z) \overline{G_{2D}^{t_{0}}}(\rho_{0}) \; dx dz + \varepsilon
  \label{eq:worst_case_total}
\end{align}
where $\varepsilon$ is some error that can be made arbitrarily small in magnitude by increasing the size of $N$, and $\overline{G_{2D}^{t_{0}}}(\rho_{0})$ is defined by
\begin{equation}
  \label{eq:Greens_priors_included}
  \overline{G_{2D}^{t_{0}}}(x,z) = G_{2D}(x,z) + \frac{i}{2}H_{0}^{(1)}(\beta \rho, c_{n}t_{0}/\rho)
\end{equation}
Hence, the size of the error term in the Green's function in equation (\ref{eq:Arrival_greens_2D}) effectively doubles when we consider the worst-case contributions of previous inputs, in addition to the current input.

\subsection*{Bounds on Operating Frequency}
\label{sec:bounds-oper-freq}

We now pose the question of how long the tick length $t_{0}$ needs to be to ensure that time-based errors in the Green's function do not cause it to differ substantially from the ideal $G_{2D}$. To do this, we enforce that $t_{0}$ be sufficiently large that
the total magnitude of the time-based error never exceeds the usual space-based error in approximating $G_{2D}$ by a complex exponential. To achieve this, we adopt asymptotic bounded approximations for $G_{2D}$ and the incomplete Hankel function, detailed in the Supplementary Information. Using these representations, we ensure that the error term in equations \eqref{eq:Arrival_greens_2D} and \eqref{eq:Greens_priors_included} is bounded in magnitude by the size of the error in approximating $G_{2D}$ by a complex exponential. This yields a minimum clock-tick duration $t_{0}$ that maintains the required error conditions. In approximate form,
\begin{equation}
  \label{eq:approximate_t_0}
  \min(t_{0}) \approx \frac{16K\sqrt{2}}{\sqrt{\pi}}\frac{n_{e}\rho\sqrt{kn_{e} \rho}}{c}
\end{equation}
where $K$ takes value $2$ if the residues of prior inputs are included in the derivation and $1$ if previous residues are excluded.
\begin{figure*}[t]
  \centering
\includegraphics[width=\textwidth]{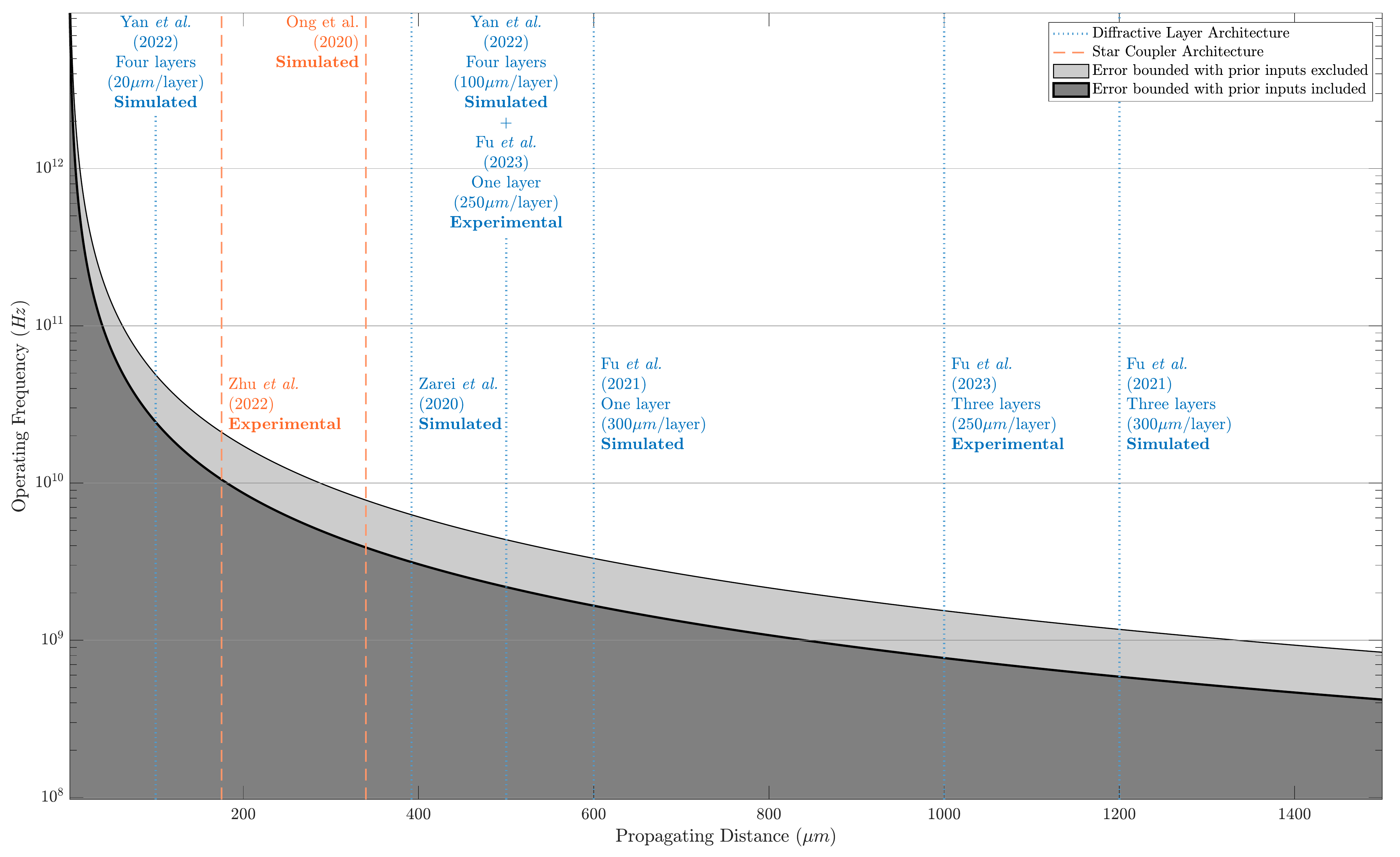}
  \caption{
    Operating frequency plotted against propagation distance $\rho$ for a 1550nm vacuum wavelength light source, and a silicon-core, silica-cladding, $0.22\mu m$ core height slab waveguide, yielding $n_{e}=2.84$. Lines corresponding to the propagation distances in recent experimental and simulated realisations of star couplers and layered diffractive architectures for optical computing applications are plotted \protect \cite{ongPhotonicConvolutionalNeural2020,zareiIntegratedPhotonicNeural2020,fuOnchipPhotonicDiffractive2021,fuPhotonicMachineLearning2023,zhuSpaceefficientOpticalComputing2022,yanAllopticalGraphRepresentation2022a}. For the layered diffractive architectures, the number of layers refers to the number of lines of metamaterials within the D2NN-type design -- the propagation distance between adjacent layers according to the published results is labelled.}
  \label{fig:plot}
\end{figure*}

From the clock tick $t_{0}$ we can obtain a maximum operating frequency as $1/t_{0}$, which we plot against $\rho$ for values of $k$ and $n_{e}$ observed in the literature, in Figure \Ref{fig:plot}. More precise bounds than the approximate representation in \eqref{eq:approximate_t_0}, given as roots of quartic equations, and detailed in the Supplementary Information, are used to delineate the shaded regions in the figure.

\section*{Discussion}
\label{sec:discussion}

It is observed that 2D diffractive photonic computing systems can be run with tolerable accuracy at rates up to at least the GHz frequency range. At frequencies in excess of the shaded bounded error regions of Figure \ref{fig:plot}, acceptable errors may still be obtained, but time-based effects will start to dominate the overall sources of error and may require greater care in applications. These operating frequencies are already within reach.  Fu \emph{et al.} found that classification accuracies began to suffer in high-operating-frequency on-chip D2NN implementations, which may in part be attributable to operating in regions of the plot in Figure \ref{fig:plot} well above the shaded boundary, without explicit measures to account for the time-based error, with 100GHz photodetection frequencies stated \cite{fuPhotonicMachineLearning2023}. Systems built upon shorter diffraction distances can be expected to be more robust at higher operating frequencies, with a higher ceiling before the frequencies cross the boundary of the respective shaded regions. Conversely, shorter diffraction distances result in larger spatial sources of error in approximating $G_{2D}$ by a complex exponential. Hence, trade-offs emerge between temporal and spatial sources of error as the propagation distance is modified in a star coupler or any design relying on a Fresnel model of propagation.

The case of one of the layered diffractive architecture due to Yan \emph{et al.} included in Figure \ref{fig:plot} differs from that of a star coupler based design in that its proper functioning does not rely on an approximation of $G_{2D}$ by a complex exponential \cite{yanAllopticalGraphRepresentation2022a}. Instead, the cited work used angular spectrum methods in the design optimisation process, which nonetheless \emph{do} ignore the diffusive nature of 2D propagation. As such, although the boundary of the respective shaded regions does not have the practical meaning of equalising spatial and temporal error contributions, it illustrates the fact that designs with longer propagation distances will be less robust to diffusive effects at high operating frequencies. Other layered diffractive architectures use a Fresnel model of diffraction in their optimisation processes \cite{fuOnchipPhotonicDiffractive2021,fuPhotonicMachineLearning2023}. By contrast with the angular spectrum method, this Fresnel approach can be shown to be representing $G_{2D}$ by a complex exponential, such that the discussion of temporal versus spatial sources of errors goes ahead as it did in the case of star coupler based designs. In another work, we show that the angular spectrum is indeed a Fourier-domain representation of a direct (time-independent) solution to the Helmholtz equation \cite{wetherfieldPlanarFourierOptics2023}.

To conclude, in this paper, we have used a time-aware analytical approach to investigate the influence of diffusive wave propagation on the performance of 2D diffractive Fourier optical systems. By drawing out a time-dependent error term in the Green's function, we are able to quantify the divergence from the ideal Green's function that it introduces in terms of asymptotic expansions. From here, we are able to explicitly plot a pair of bounds on the timing constraints of a diffractive photonic system, in terms of its wavelength and effective refractive index parameters, that depends on propagation distance. For one bound, we factor in the diffusive residue left by prior system inputs, while for the other, we disregard these prior inputs. Together, the bounds reveal a range of operating speeds at which a system can run without introducing dominant temporal sources of error. As well as providing a framework for understanding sources of error at higher operating frequencies, the analysis also provides a range of speeds at which the time-independent models of wave propagation can be deemed sufficiently accurate for a given application (such that the diffusive nature of 2D wave propagation can be overlooked). As the operating speeds of integrated diffractive photonic circuits begin to compete with conventional silicon electronics, we intend that the analysis in this work helps to explain new temporal sources of errors that may emerge. At lower operating speeds, we have shown that time-independent models only introduce an insignificant error, and hence remain a useful first approximation.

\section*{Acknowledgements}

We thank X. Lin, H. H. Zhu, L. C. Kwek, J. R. Ong and C. C. Ooi for clarifying correspondence on their work. We also thank T. Albrow-Owen for helpful comments on this manuscript.

This work was supported by the Richard Norman Scholarship grant for the Department of Engineering, University of Cambridge. For the purpose of open access, the authors have applied a Creative Commons Attribution (CC BY) licence to any Author Accepted Manuscript version arising.

\begin{spacing}{1.15}
\bibliographystyle{naturemag}
\bibliography{bibliography.bib}
\end{spacing}

\end{document}


\maketitle

\section*{Supplementary Information}
\label{sec:suppl-inform}

\localtableofcontents

\section{Asymptotic Expansions}
\label{sec:asympt-expans}

In order to solve for bounds on $t_{0}$ and $1/t_{0}$, as introduced in the main text, we use asymptotic expansions for complete and incomplete Hankel functions, which are valid for large argument (corresponding to propagation distance). The following asymptotic expansion for the complete Hankel function is standard (see pages 196-197 of Watson's classic treatise \cite{watson}).

\begin{equation}
  \label{eq:Hankel_bound}
  H_{0}^{(1)}(\vert x \vert ) = \sqrt{\frac{2}{i\pi\vert x \vert}}\exp(i\vert x \vert - \pi / 4) + R(\vert x \vert)
\end{equation}
Here, $R$ is bounded for real positive argument:
\begin{equation}
  \label{eq:bound_on_R}
  \lvert R(\lvert x \rvert) \rvert \le \frac{1}{8\lvert x \rvert} \sqrt{\frac{2}{\pi\lvert x \rvert}}
\end{equation}
Second, we prove the following bounds for our defined incomplete Hankel function:
\begin{equation}
  \label{eq:asymptotic_expansion}
  H_{0}^{(1)}(\lvert x \rvert, T) = \frac{2\exp(i\lvert x \rvert T)}{i\pi\lvert x \rvert \sqrt{T^{2} - 1}} + R(\lvert x \rvert, T)
\end{equation}
where the remainder term $R(\lvert x \rvert, T)$ is bounded as follows:
\begin{equation}
  \label{eq:bound_on_remainder}
  \vert R(\lvert x \rvert, T) \lvert \le \frac{4}{\pi\lvert x \rvert^{2}}\frac{T}{(T^{2} - 1)^{3/2}}
\end{equation}
and, more coarsely,
\begin{equation}
  \label{eq:bound_on_incomplete_hankel}
  \left \lvert H_{0}^{(1)}(\lvert x \rvert, T) \right \rvert \le \frac{4}{\pi}\frac{1}{\lvert x \rvert \sqrt{T^{2} - 1}}
\end{equation}
Both of these bounds follow from the following theorem.\footnote{A similar theorem appears in Wong's classic text \protect\cite{wong} (pages 16-17).}

\noindent\hrulefill

\begin{theorem}
 Let $f(t)$ be continuous and infinitely differentiable in the region $[T,\infty]$ such that $f^{(N)}(t)~=~0$ for all $N$ as $t \to \infty$. Then,
\begin{equation}
  \label{eq:thm}
  \int_{T}^{\infty} f(t)\exp(i\lvert x \rvert t) \; dt = \sum_{n=0}^{N-1}\left ( \frac{i}{\lvert x \rvert }\right )^{n + 1}f^{(n)}(T)\exp(i\lvert x \rvert T) + \varepsilon_{N}(\lvert x \rvert, T )
\end{equation}
where the error term $\varepsilon_{N}$ is bounded as follows:
\begin{equation}
  \label{eq:thm_error_bound}
  \lvert \varepsilon_{N}(\lvert x \rvert, T) \rvert \le \left ( \frac{1}{\lvert x \rvert }\right )^{N + 1} \left ( \left \lvert f^{(N)}(T) \right \rvert +  \int_{T}^{\infty}\left \lvert f^{(N+1)}(t) \right \rvert\; dt  \right )
\end{equation}
If, in addition, for each $n$, $f^{(n)}(t) \ge 0$ or $f^{(n)}(t) \le 0$ for all $t \in [T, \infty]$, the error bound becomes:
\begin{equation}
  \label{eq:thm_error_bound_2}
  \lvert \varepsilon_{N}(\lvert x \rvert, T) \rvert \le 2 \left ( \frac{1}{\lvert x \rvert }\right )^{N + 1} \left ( \left \lvert f^{(N)}(T) \right \rvert \right )
\end{equation}
\end{theorem}

\noindent\hrulefill

Hence, setting $f(t)=2(t^{2} - 1)^{1/2}/(i\pi)$, and $T>1$, we have
\begin{equation}
  \label{eq:specific_asymptotic_expansion}
  H_{0}^{(1)}(\lvert x \rvert, T) = \frac{2\exp(i\lvert x \rvert T)}{i\pi}\sum_{n=0}^{N-1}\left ( \frac{i}{\lvert x \rvert }\right )^{n + 1}\frac{d^{n}}{dT^{n}}\left (\frac{1}{\sqrt{T^{2} - 1}}\right ) + R_{N}(\lvert x \rvert )
\end{equation}
with $R_{N}$ satisfying:
\begin{equation}
  \label{eq:specific_asymptotic_error_bound}
  \lvert R_{N}(\lvert x \rvert) \rvert \le \frac{4}{\pi} \left ( \frac{1}{\lvert x \rvert }\right )^{N + 1} \left ( \left \lvert \frac{d ^{N}}{dT^{N}}\frac{1}{\sqrt{T^{2} - 1}} \right \rvert \right )
\end{equation}

Choosing $N = 0$ and $1$, we obtain the specific asymptotic bounds given in equations \eqref{eq:bound_on_incomplete_hankel} and \eqref{eq:bound_on_remainder} respectively. For large enough values of $x$ (or $\rho$ in the context of the main text), larger values of $N$ may yield tighter bounds on $H_{0}^{(1)}(x, T)$.

\begin{proof}[Proof of the theorem]
Integrating by parts $N$ times yields
\begin{equation}
  \label{eq:integration_by_parts}
\int_{T}^{\infty}f(t)\exp(i\vert x \vert t) = \sum_{n=0}^{N-1} \left (\frac{i}{\lvert x \rvert }\right  )^{n+1} \exp(i\lvert x \rvert T) + \varepsilon_{N}(\lvert x \rvert, T)
\end{equation}
where
\begin{equation}
  \label{eq:epsilon}
  \varepsilon_{N}(\lvert x \rvert, T) =  \int_{T}^{\infty} \left (\frac{i}{\lvert x \rvert}\right )^{N}f^{(N)}(t) \exp(i\lvert x \rvert t)  dt
\end{equation}
A further integration by parts gives
\begin{equation}
  \label{eq:further_parts}
  \varepsilon_{N}(\lvert x \rvert, T) =  \left (\frac{i}{\lvert x \rvert}\right )^{N+1}\left [f^{(N)}(T) \exp(i\lvert x \rvert T) +
\int_{T}^{\infty}  f^{(N+1)}(t) \exp(i\lvert x \rvert t) \; dt \right ]
\end{equation}
Hence, by the triangle inequality, we have
\begin{equation}
  \label{eq:triangle}
  \lvert \varepsilon_{N}(\lvert x \rvert, T) \rvert \le  \left (\frac{1}{\lvert x \rvert}\right )^{N+1}\left [\left \lvert f^{(N)}(T) \right \rvert +
\int_{T}^{\infty}\left |  f^{(N+1)}(t) \right | \; dt \right ]
\end{equation}

The final condition in the theorem provides
\begin{align}
  \label{eq:final_condition_outcome}
  \int_{T}^{\infty}\left \lvert f^{(N+1)}(t) \right \rvert \; dt &= \left \lvert\int_{T}^{\infty}  f^{(N+1)}(t)  \; dt\right \rvert \\
  &= \left \lvert f^{(N)}(T) \right \rvert
\end{align}
which completes the final part of the proof.

\end{proof}

\pagebreak
\section{Solving for Bounds}
\label{sec:solving-bounds}

As stated in the main text, the sought criterion on $t_{0}$ and $1/t_{0}$ can be achieved by ensuring that the error incurred from temporal effects is no larger in magnitude than the error accepted from approximating $G_{2D}$ by a complex exponential. Armed with the asymptotic bounds from the previous section (equations (\ref{eq:Hankel_bound}-\ref{eq:bound_on_remainder})), we wish to ensure:
\begin{equation}
  \label{eq:bounding_condition_no_priors}
  \lvert H_{0}^{(1)}(\beta \rho, c_{n}t_{0}/\rho) \rvert \le \lvert R(\beta \rho)  \rvert
\end{equation}
in the case that we \emph{disregard} prior inputs $F_{j}$, per the main text, or,
\begin{equation}
  \label{eq:bounding_condition}
  2\lvert H_{0}^{(1)}(\beta \rho, c_{n}t_{0}/\rho) \rvert \le  \lvert R(\beta \rho)  \rvert
\end{equation}
in the case that we \emph{include} prior inputs $F_{j}$.

Hence, using equations (\ref{eq:asymptotic_expansion}) and (\ref{eq:bound_on_remainder}), we set
\begin{equation}
  \label{eq:set_bound}
  K\cdot\frac{2}{\pi}\left (\frac{1}{X}\frac{1}{\sqrt{T^{2} - 1}} + \frac{2T}{X(T^{2} - 1)^{2}}\right) \le \sqrt{\frac{2}{\pi X}} \frac{1}{8X}
\end{equation}
with $X = \beta \rho$ and $T = c_{n}t_{0} /\rho$, and where $K$ is $1$ or $2$, depending on whether we solve for case \eqref{eq:bounding_condition_no_priors} or \eqref{eq:bounding_condition}.
To find suitable $t_0$, we can seek the slightly relaxed bound
\begin{equation}
  \label{eq:mildly_relaxed_bound}
  \frac{2K}{\pi}\left (\frac{(T^{2} -1)^{3/2}}{X} + \frac{2T}{X}\right)T < \sqrt{\frac{2}{\pi X}} \frac{1}{8X}(T^{2} - 1)^{2}
\end{equation}
and find roots of the resulting quartic in $t_{0}$ obtained by setting left- and right-hand-sides equal. Of the four roots, the one with physical significance is the one close to the positive root of the following quadratic (where the coarser bound \eqref{eq:bound_on_incomplete_hankel} on $\lvert H_{0}^{(1)}(X, T) \rvert $ has been used):
\begin{equation}
  \label{eq:simpler_bound_condition}
  \frac{4K}{\pi}  \frac{T}{X}  - \sqrt{\frac{2}{\pi X}}\frac{1}{8X}(T^{2} - 1) = 0
\end{equation}
yielding
\begin{equation}
  \label{eq:approximate_t_0}
  \min(t_{0}) \approx \frac{16K\sqrt{2}}{\sqrt{\pi}}\frac{n_{e}\rho\sqrt{kn_{e} \rho}}{c}
\end{equation}
where $n_{e}$ is the effective refractive index of the wave in the medium, $k$ is the free-space wavenumber and $c$ is the speed of light in a vacuum, as stated in the main text. For tighter bounds on $t_{0}$, higher order bounded asymptotic approximations of $H_{0}^{(1)}(X,T)$, as given in the previous section, can be used.

\pagebreak
\section{Procedure for Generating Main Plot}
\label{sec:proc-gener-plot}

The following procedure outlines the method for generating the main plot. Perform the following steps for $K=1$ and $K = 2$, where $c$ is the speed of light in a vacuum, $c_{n} = c/n_e$, $n_e =2.84$, $\beta = kn_{e}$, $k = 2\pi / \lambda$ and $\lambda = 1550\times 10^{-9}$:

\begin{algorithmic}
  \algblockdefx{With}{End}[1]{\textbf{With} #1:}{\textbf{End}}
  \State $\textrm{TimeQuartic}(T, X, K) \gets
  \displaystyle \frac{2K}{\pi}\left (\frac{(T^{2} -1)^{3/2}}{X} + \frac{2T}{X}\right)T - \sqrt{\frac{2}{\pi X}} \frac{1}{8X}(T^{2} - 1)^{2}$
  \State $\textrm{FreqQuartic}(F, X, K) \gets \textbf{Simplify }F^{4} \times\textrm{TimeQuartic}(1/F, X, K)$
  \State $\textrm{TimeQuadratic}(T, X, K) \gets
  \displaystyle
  \frac{4K}{\pi}  \frac{T}{X}  - \sqrt{\frac{2}{\pi X}}\frac{1}{8X}(T^{2} - 1) $
  \State $\textrm{FreqQuadratic}(F, X, K) \gets
  \displaystyle
  \textbf{Simplify } F^{2} \times \textrm{TimeQuadratic}(1/F, X, K)$
  \State QuarticRoots[4] $\gets \textbf{Solve }\textrm{FreqQuartic}(\rho f/c_{n}, \beta \rho, K) = 0 \textrm{ in terms of } f$
  \State QuadraticRoots[2] $\gets \textbf{Solve }\textrm{FreqQuadratic}(\rho f/c_{n}, \beta \rho, K) = 0 \textrm{ in terms of } f$
  \With{$\rho \in [1\times 10^{-6}, 1.5\times 10^{-4}]$}
    \State \textrm{PhysicalQuadraticRoot} $\gets \textbf{Select }$ maximum root in QuadraticRoots
    \State \textrm{PhysicalQuarticRoot} $\gets \textbf{Select }$ root in QuarticRoots with same order of magnitude as PhysicalQuadraticRoot
    \State $\textbf{Plot }\textrm{ PhysicalQuadraticRoot against } \rho$
  \End
\end{algorithmic}

\section*{Rights Statement}

This work was supported by the Richard Norman Scholarship grant for the Department of Engineering, University of Cambridge.
For the purpose of open access, the authors have applied a Creative Commons Attribution (CC BY) licence to any Author Accepted Manuscript version arising.
